\documentclass[final]{aipproc}
\layoutstyle{6x9}

\begin{document}

\title{Testing the Standard Model with Kaon Decays}


\author{Matthew Moulson}
  {address={Laboratori Nazionali di Frascati dell'INFN, Via E. Fermi, 40, 
00044 Frascati RM, Italy}
}

\begin{abstract}
During the last few years, 
new experimental and theoretical results have allowed ever-more stringent 
tests of the Standard Model to be performed using kaon decays. This overview 
of recent progress includes an updated evaluation of the CKM matrix element 
$V_{us}$. Tests of CKM unitarity and gauge universality and lepton
universality tests with $K_{l2}$ and $K_{l3}$ decays are discussed. 
\end{abstract}

\maketitle

\section{$V_{us}$ from $K_{\ell3}$ and $K_{\ell2}$ decays; related tests}
If the couplings of the $W$ to quarks and leptons are indeed specified 
by a single gauge coupling, then for universality to be observed as the 
equivalence of the Fermi constant $G_F$ as measured in muon and hadron decays, 
the CKM matrix must be unitary. Currently, the most stringent test of 
CKM unitarity is obtained from the first-row relation
$|V_{ud}|^2 + |V_{us}|^2 + |V_{ub}|^2 = 1 + \Delta_{\mathrm{CKM}}$.
During the period spanning 2003 to 2010, a wealth of new measurements
of $K_{\ell3}$ and $K_{\ell2}$ decays and
steady theoretical progress made possible precision tests of the Standard 
Model (SM) based on this relation. In a 2010 evaluation of $|V_{us}|$, the 
FlaviaNet Working Group on Kaon Decays set bounds on $\Delta_{\mathrm{CKM}}$
at the level of 0.1\%~\cite{FlaviaNet+10:Vus}, which translate into bounds 
on the effective scale of new physics on the order of 10~TeV~\cite{CJGA10:eff}.
Since 2010, there have been a few significant new measurements and some 
important theoretical developments. Among the latter, advances in algorithmic 
sophistication and computing power are leading to more and better lattice
QCD estimates of the hadronic constants $f_+(0)$ and $f_K/f_\pi$, which 
enter into the determination of $|V_{us}|$ from $K_{\ell 3}$ and $K_{\mu2}$ 
decays, respectively. In addition, two groups working on the classification 
and averaging of results from lattice QCD~\cite{FLAG+11:review,LLVdW+10:fits}
have joined their efforts, forming the newly formed Flavor Lattice Average 
Group (FLAG-2) to provide recommended values of these 
constants~\cite{Col12:Lattice}. 

The experimental inputs for the determination of $|V_{us}|$ from $K_{\ell3}$
decays are the rates and form factors for the decays of both charged and 
neutral kaons. There have been no new branching ratio (BR) measurements 
since the 2010 review. On the other hand, both the KLOE and KTeV 
collaborations have new measurements of the $K_S$ lifetime, $\tau_{K_S}$. 
The KLOE measurement~\cite{KLOE+11:KSlife} is based on the vertex distribution 
for $K_S\to\pi^+\pi^-$ decays in $e^+e^-\to\phi\to K_SK_L$ events.
The new KTeV result~\cite{KTeV+11:epspr} comes from a comprehensive 
reanalysis of the $\pi\pi$ vertex distributions in the experiment's 
regenerator beam. This analysis can be performed with or without assuming 
$CPT$ symmetry. The present update makes use of the new KTeV values for
$\tau_{K_S}$ and $\mathrm{Re}\,\varepsilon'/\varepsilon$ 
(which enters the fit for the $K_L$ rates by providing an effective 
measurement of $\mathrm{BR}(\pi^0\pi^0)/\mathrm{BR}(\pi^+\pi^-)$)
obtained without the $CPT$ assumption. 
The largest effect of these updates is to reduce the uncertainty on
$\tau_{K_S}$, the value of which changes from $89.59(6)$~ps
\cite{FlaviaNet+10:Vus} to $89.58(4)$~ps.

\begin{figure}
\includegraphics[height=0.28\textheight]{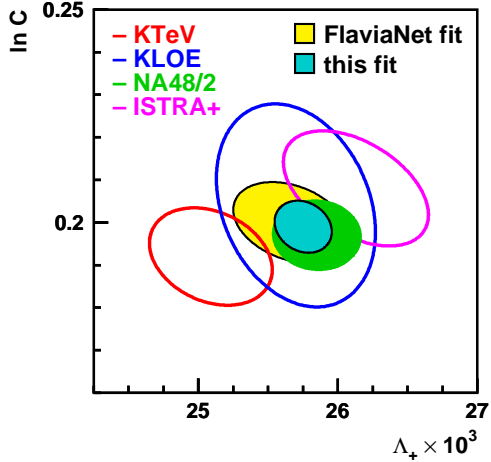}
\caption{$1\sigma$ confidence contours for form factor parameters 
($K_{e3}$-$K_{\mu3}$ averages) from dispersive fits, for different experiments. 
The NA48/2 result was converted by the author from the experiment's polynomial 
fit results. The FlaviaNet 2010 average and the new average, with the NA48/2
result included, are also shown.}
\label{fig:ffcomp}
\end{figure} 
The NA48/2 experiment has recently released preliminary results for the 
form factors for both $K^\pm_{e3}$ and $K^\pm_{\mu3}$ decays~\cite{HH:Moriond12}.
At the moment, the fits are performed using one of two parameterizations,
\begin{itemize}
\item polynomial: $(\lambda'_+, \lambda''_+)$ for $K_{e3}$, 
$(\lambda'_+, \lambda''_+, \lambda_0)$ for $K_{\mu3}$, or
\item polar: $M_V$ for $K_{e3}$, $(M_V, M_S)$ for $K_{\mu3}$.
\end{itemize}
The FlaviaNet Working Group uses the dispersive parameterizations 
of~\cite{B+09:disp} because of the advantages described 
in~\cite{FlaviaNet+10:Vus}. NA48/2 has not yet performed dispersive fits.
However, it is possible to fit approximately equivalent values
of $\Lambda_+$ and $\mathrm{ln}\,C$ to the NA48/2 measurements of
$(\lambda'_+, \lambda''_+, \lambda_0)$ using the expressions in the appendix 
of~\cite{B+09:disp} and the observation that 
$\lambda_0 \approx \lambda_0' + 3.5\lambda_0''$ \cite{KLOE+07:Km3FF}.
This is important because it helps to resolve a controversy: the older
measurements of the $K_{\mu3}$ form factors for $K_L$ decays from 
NA48~\cite{NA48+07:m3FF} are in such strong disagreement with the other 
existing measurements that they have been excluded from the FlaviaNet 
averages~\cite{FlaviaNet+10:Vus}. 
The new NA48/2 measurements, on the other hand, are in good agreement with 
other measurements, as seen from Fig.~\ref{fig:ffcomp}. Including the new 
NA48/2 results, appropriately converted for the current purposes, the 
dispersive average becomes $\Lambda_+ = (25.75\pm0.36)\times10^{-3}$, 
$\mathrm{ln}\,C = 0.1895(70)$, with $\rho = -0.202$ and $P(\chi^2) = 55\%$.
The central values of the phase space integrals barely change with this 
inclusion; the uncertainties are reduced by 30\%. 

For all of the above efforts, however, the value of and uncertainty on 
$|V_{us}|\,f_+(0)$ is essentially unchanged. This is because the new results
are nicely consistent with the older averages, and neither the 
$K_S$ lifetime nor the phase space integrals were significant contributors 
to the overall experimental uncertainty. 
The updated five-channel ($K_{L,e3}$, $K_{L,\mu3}$, $K_{S,e3}$, $K^\pm_{e3}$,
$K^\pm_{\mu3}$) average is $|V_{us}|\,f_+(0) = 0.2163(5)$ with  
$P(\chi^2) = 93\%$. The ratio of the values for $|V_{us}\,f_+(0)|^2$
obtained from $K_{\mu3}$ and $K_{e3}$ decays is 1.002(5). This ratio is 
proportional to $(g_\mu/g_e)^2$; the result confirms the universality of lepton couplings. Comparison of the results obtained with $K^+$ and $K_L$, $K_S$ 
decays confirms the accuracy of the correction for strong isospin breaking
$\Delta_{SU(2)}$. The correction used in the analysis is
$\Delta_{SU(2)} = (2.9\pm0.4)\%$ \cite{KN08:Kl3FF}, 
while perfect equality of the experimental results would require 
$\Delta_{SU(2)} = (2.73\pm0.41)\%$. The uncertainty 
on the theoretical value is one of the largest contributions to the
uncertainty on $|V_{us}|\,f_+(0)$ from $K^\pm$ decays.

The currently recommended FLAG-2 value~\cite{Col12:Lattice} 
for $f_+(0)$ from three-flavor lattice QCD is that obtained by 
RBC/UKQCD~\cite{UKRBC+10:f0}. For the present 
analysis, the uncertainty is symmetrized, giving $f_+(0) = 0.959(5)$
and thus $|V_{us}| = 0.2254(13)$. Importantly, the test of CKM unitarity also 
requires a value for $|V_{ud}|$. The most recent definitive survey of 
experimental data on $0^+\to0^+$ $\beta$ decays is that of~\cite{HT09:VudSFT}, 
which gives $|V_{ud}| = 0.97425(22)$. This, together with the value of 
$|V_{us}|$ from $K_{\ell3}$, yields $\Delta_\mathrm{CKM} = 0.0000(8)$, 
demonstrating perfect agreement with unitarity.

Up to kinematic factors and radiative corrections, the ratio of 
the (inner-bremsstrahlung inclusive) rates for $K_{\mu2}$ and $\pi_{\mu2}$ 
decays provides the quantity $|V_{us}/V_{ud}|\times f_K/f_\pi$. 
The FlaviaNet fit for the $K^\pm$ BRs, which provides the values
for $\mathrm{BR}(K^\pm_{\mu2})$ and $\tau_{K^\pm}$, is unchanged since the 2010 
review. There are some important changes in the theoretical inputs. 
The most recent FLAG-2 average of the four complete and published 
determinations of $f_K/f_\pi$ in three-flavor lattice QCD is
$f_K/f_\pi = 1.193(5)$~\cite{Col12:Lattice}, only 
slightly changed from the situation in 2010 by the addition of a single study.
More significant is a new calculation of the corrections to the 
ratio $\Gamma(K^\pm_{\mu2})/\Gamma(\pi^\pm_{\mu2})$ for isospin 
breaking~\cite{CN11:Kl3IB}, which for the first time 
takes into account the effects of strong isosping breaking, in addition to the 
long-distance electromagnetic corrections. While the magnitude of the total 
correction is nearly doubled, the contribution to the uncertainty from 
the correction is increased by only about 20\%. The resulting value for
$|V_{us}/V_{ud}|$ is 0.2317(11). 

The values of $|V_{ud}|$ from $0^+\to0^+$ $\beta$ decays, $|V_{us}|$ from 
$K_{\ell3}$ decays, and $|V_{us}/V_{ud}|$ from $K_{\mu2}$ decays can be combined
in a single fit to increase the sensitivity of the unitarity test.
The unconstrained fit does not change the 
input value of $|V_{ud}|$ and gives $|V_{us}| = 0.2256(8)$. 
This gives $\Delta_\mathrm{CKM} = +0.0001(6)$, once again in perfect agreement 
with unitarity. Using a model-indepenent, effective-theory approach,
the authors of~\cite{CJGA10:eff} show that the effective scale for 
corresponding contributions from new physics with approximate flavor symmetry
is about 10~TeV.  

The corresponding results in 2010 were $|V_{us}| = 0.2253(9)$ 
and $\Delta_\mathrm{CKM} = -0.0001(6)$ \cite{FlaviaNet+10:Vus}. 
The new measurements of $\tau_{K_S}$ and 
the $K_{\ell3}$ form factor parameters have virtually no effect on the 
final result. While there is a small effect from the inclusion of the 
new lattice estimate for $f_K/f_\pi$ in the FLAG-2 average, the largest single
influence is from the new correction to the ratio 
$\Gamma(K_{\mu2})/\Gamma(\pi_{\mu2})$ for the effects of strong isospin 
breaking. This underscores a simple fact: so much work has been done to 
increase the precision of the experimental inputs to $|V_{us}|$ that, for
the moment at least, further experimental progress is difficult. 
The measurements that offer the most room for improvement are the BRs 
for the $K_{\ell3}$ decays of the $K_S$ and of the $K^\pm$, and in the case 
of $K^\pm$, to be useful, better BR measurements would also 
require a more precise theoretical estimate for $\Delta_{SU(2)}$.
But while the experimental quantities $|V_{us}|f_+(0)$ and 
$|V_{us}/V_{ud}|\times f_K/f_\pi$ have been measured to within about 
0.2\%, the precision of the unitarity test is currently determined by the 
uncertainties on the lattice results for $f_+(0)$ and $f_K/f_\pi$, 
which are at the level of 0.5\%. Thus, at the moment, the lattice offers
the most certain prospects for further improvement. Results for $f_+(0)$ 
and $f_K/f_\pi$ with precison at the level of 0.2\% may be available as 
early as 2014, and continued progess is expected thereafter~\cite{VdW12:CIPANP}.

\section{$K_{\ell2}$ decays and lepton universality}

The $K_{e2}$ decay ($K\to e\nu$) is strongly helicity suppressed. 
The ratio $R_K$ of the widths (inclusive of internal bremsstrahlung) 
for $K_{e2}$ to $K_{\mu2}$ decays depends only on kinematic factors and
radiative corrections---the form factor $f_K$ cancels from 
the ratio, and the uncertainty on the expected value is very small:
$R_K = 2.477(1) \times 10^{-5}$ \cite{CR07:Kp2}. 
These features---a BR suppressed in the SM
and a precisely known SM rate---make the decay interesting in the 
precision-based search for evidence of new physics. A specific motivation
for new measurements of $R_K$ is that in the Minimal Supersymmetric Standard
Model (MSSM) with $R$ parity, there may be a percent-level contribution to 
the $K_{e2}$ rate from the $H^+$-mediated amplitude with lepton-flavor
violation in the one-loop effective coupling at the $H^+\to e^+\nu$ vertex
(i.e., with $\nu = \nu_\tau$), depending on the values of $\tan\beta$, 
$m_{H^+}$,and the mass insertion $|\Delta^{31}_R|$ in the loop~\cite{MPP06:Ke2}.

The first modern measurement of $R_K$ was performed with the KLOE 
experiment~\cite{KLOE+09:Ke2}. The $K^\pm\to e^\pm$ decay vertices were 
reconstructed in the drift chamber with tight quality cuts. The event 
selection was refined using the missing mass at the vertex 
($M_\ell^2 = m_\nu^2 = 0$ for signal events) and a neural network for $e/\mu$
separation in the electromagnetic calorimeter. The event counting was 
performed by a two-dimensional likelihood fit in the plane of $M_\ell^2$ vs.\ 
the neural network output. On the basis of about 14k events, KLOE measured 
$R_K=2.493(25)_\mathrm{stat}(19)_\mathrm{syst}$, i.e., with an uncertainty of 
about 1.2\%. KLOE was also able to isolate $K_{e2\gamma}$ events with 
structure-dependent (SD+) radiation well enough to fit the $E_\gamma$ 
spectrum to determine the parameters for the $\mathcal{O}(p^4)$ 
chiral-perturbation-theory representation of the form factor.

The NA62 experiment, a successor to NA48 the goal of which is to measure the
rare $K^+\to\pi^+\nu\bar{\nu}$ decay, benefits from much higher statistics from
the 75-GeV kaon beam and easier $e/\mu$ separation at high energies with
the NA48 liquid krypton calorimeter (LKr). In 2007, NA62 collected nearly 
150k $K_{e2}$ decays in a dedicated run. For the analysis, a cut on the 
$E/p$ ratio for decay tracks (with $E$ measured in the LKr and $p$ 
reconstructed in the dipole spectrometer) provides $\mu$ rejection at 
the level of $10^{-6}$. The probability for muons to lose enough energy 
in the LKr via catastrophic bremsstrahlung to be mistaken for electrons 
is estimated using data from runs in which part of the LKr acceptance
was blocked to electrons with a thick lead shield. Event counting is 
performed by fitting the $M_\ell^2$ distributions. In 2011, NA62 published 
a result based on a 40\% subsample of the data collected: 
$R_K = 2.487(13)\times 10^{-5}$ \cite{NA62+11:RK}. 
The full sample has now been analyzed and the final result made public:
$R_K = 2.488(7)_\mathrm{stat}(7)_\mathrm{syst}\times 10^{-5}$
\cite{NA62+12:RK_CIPANP}. This result has 0.4\% precision and is in agreement 
with the SM prediction. For the subtraction of the background from the 
SD+ component of the radiative channel $K_{e2\gamma}$, the NA62 measurement
makes use of the KLOE measurement mentioned above. NA62 has performed its
own measurement of the SD+ component in the works and preliminary results have
been shown~\cite{NA62+12:Ke2g_CIPANP}.

\begin{figure}
\includegraphics[height=0.28\textheight]{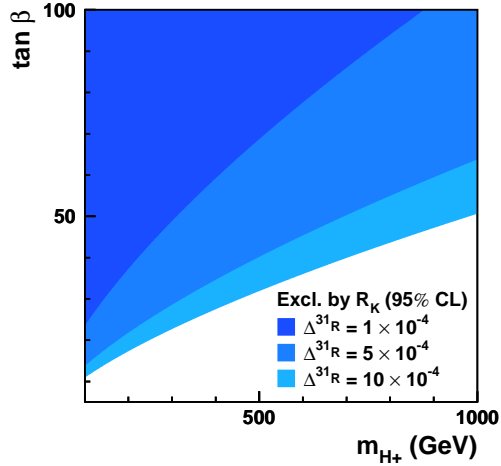}
\caption{Regions in the $(m_{H^+}, \tan\beta)$ plane excluded at 95\% CL 
by the world average value for $R_K$ (including the newest NA62 result), 
for different values assumed for the mass insertion $|\Delta^{31}_R|$
in the MSSM.}
\label{fig:higgs_rk}
\end{figure} 
The new NA62 result essentially determines the new world average:
$R_K = 2.488(9) \times 10^{-5}$. The regions in the $(m_{H^+}, \tan\beta)$ 
plane excluded at 95\% CL by the agreement of this value with the SM 
prediction are shown in Fig.~\ref{fig:higgs_rk} for different values 
assumed for $|\Delta^{31}_R|$ in the MSSM. 

\bibliographystyle{aipproc}
\bibliography{214_Moulson}

\begin{thebibliography}{22}
\expandafter\ifx\csname natexlab\endcsname\relax\def\natexlab#1{#1}\fi
\providecommand{\enquote}[1]{``#1''}
\expandafter\ifx\csname url\endcsname\relax
  \def\url#1{\texttt{#1}}\fi
\expandafter\ifx\csname urlprefix\endcsname\relax\def\urlprefix{URL }\fi
\providecommand{\eprint}[2][]{\url{#2}}

\bibitem[{FlaviaNet Kaon Working Group} et~al.(2010)]{FlaviaNet+10:Vus}
{FlaviaNet Kaon Working Group}, M.~Antonelli, et~al., \emph{Eur.\ Phys.\ J.\ C}
  \textbf{69}, 399 (2010).

\bibitem[{FLAG Working Group} et~al.(2011)]{FLAG+11:review}
{FLAG Working Group}, G.~Colangelo, et~al., \emph{Eur.\ Phys.\ J.\ C}
  \textbf{71}, 1695 (2011).

\bibitem[Laiho et~al.(2010)]{LLVdW+10:fits}
J.~Laiho, E.~Lunghi, and R.~Van~de Water, \emph{Phys.\ Rev.\ D} \textbf{81},
  034503 (2010).

\bibitem[{KLOE~Collaboration} et~al.(2011)]{KLOE+11:KSlife}
{KLOE~Collaboration}, F.~Ambrosino, et~al., \emph{Eur.\ Phys.\ J.\ C}
  \textbf{71}, 1604 (2011).

\bibitem[{KTeV~Collaboration} et~al.(2011)]{KTeV+11:epspr}
{KTeV~Collaboration}, E.~Abouzaid, et~al., \emph{Phys.\ Rev.\ D} \textbf{83},
  092001 (2011).

\bibitem[{Hita-Hochgesand for the NA48/2 Collaboration}(2012)]{HH:Moriond12}
M.~{Hita-Hochgesand for the NA48/2 Collaboration} (2012), talk at Moriond EW
  '12 conference.

\bibitem[Bernard et~al.(2009)]{B+09:disp}
V.~Bernard, et~al., \emph{Phys.\ Rev.\ D} \textbf{80}, 034034 (2009).

\bibitem[{KLOE~Collaboration} et~al.(2007)]{KLOE+07:Km3FF}
{KLOE~Collaboration}, F.~Ambrosino, et~al., \emph{JHEP} \textbf{0712}, 105
  (2007).

\bibitem[{NA48~Collaboration} et~al.(2007)]{NA48+07:m3FF}
{NA48~Collaboration}, A.~Lai, et~al., \emph{Phys.\ Lett.\ B} \textbf{647}, 341
  (2007).

\bibitem[Kastner and Neufeld(2008)]{KN08:Kl3FF}
A.~Kastner, and H.~Neufeld, \emph{Eur.\ Phys.\ J.\ C} \textbf{57}, 541 (2008).

\bibitem[Colangelo(2012)]{Col12:Lattice}
G.~Colangelo (2012), talk at Lattice '12 conference (Cairns, Australia).

\bibitem[{RBC and UKQCD~Collaborations} et~al.(2010)]{UKRBC+10:f0}
{RBC and UKQCD~Collaborations}, P.~Boyle, et~al., \emph{Eur.\ Phys.\ J.\ C}
  \textbf{69}, 159 (2010).

\bibitem[Hardy and Towner(2009)]{HT09:VudSFT}
J.~Hardy, and I.~Towner, \emph{Phys.\ Rev.\ C} \textbf{79}, 055502 (2009).

\bibitem[Cirigliano and Neufeld(2011)]{CN11:Kl3IB}
V.~Cirigliano, and H.~Neufeld, \emph{Phys.\ Lett.\ B} \textbf{700}, 7 (2011).

\bibitem[Cirigliano et~al.(2010)]{CJGA10:eff}
V.~Cirigliano, J.~Jenkins, and M.~Gonz\'alez-Alonso, \emph{Nucl.\ Phys.\ B}
  \textbf{830}, 95 (2010).

\bibitem[Van~de Water(2012)]{VdW12:CIPANP}
R.~Van~de Water (2012), talk at CIPANP '12 conference, St.\ Petersburg, FL.

\bibitem[Cirigliano and Rosell(2007)]{CR07:Kp2}
V.~Cirigliano, and I.~Rosell, \emph{Phys.\ Rev.\ Lett} \textbf{99}, 231801
  (2007).

\bibitem[Masiero et~al.(2006)]{MPP06:Ke2}
A.~Masiero, P.~Paradisi, and R.~Petronzio, \emph{Phys.\ Rev.\ D} \textbf{74},
  011701(R) (2006).

\bibitem[{KLOE~Collaboration} et~al.(2009)]{KLOE+09:Ke2}
{KLOE~Collaboration}, F.~Ambrosino, et~al., \emph{Eur.\ Phys.\ J.\ C}
  \textbf{64}, 627 (2009).

\bibitem[{NA62~Collaboration} et~al.(2011)]{NA62+11:RK}
{NA62~Collaboration}, C.~Lazzeroni, et~al., \emph{Phys.\ Lett.\ B}
  \textbf{698}, 105 (2011).

\bibitem[Moulson and {Balev for the NA48/2
  Collaboration}(2012)]{NA62+12:RK_CIPANP}
M.~Moulson, and S.~{Balev for the NA48/2 Collaboration} (2012), poster at
  CIPANP '12 conference, St.\ Petersburg, FL.

\bibitem[Pepe and {Romano for the NA48/2
  Collaboration}(2012)]{NA62+12:Ke2g_CIPANP}
M.~Pepe, and A.~{Romano for the NA48/2 Collaboration} (2012), poster at CIPANP
  '12 conference, St.\ Petersburg, FL.

\end{thebibliography}

\end{document}